\documentclass[manuscript,screen,nonacm]{acmart}
\acmConference[]{} 

\newcommand{\workshopname}{GenAICHI: CHI 2025 Workshop on Generative AI and HCI}
\newcommand{\licensedetails}{Licensed under a Creative Commons Attribution 4.0 International License (CC BY 4.0). Copyright remains with the author(s).}
\newcommand\extrafootertext[1]{
    \bgroup
    \renewcommand\thefootnote{\fnsymbol{footnote}}%
    \renewcommand\thempfootnote{\fnsymbol{mpfootnote}}%
    \footnotetext[0]{#1}%
    \egroup
}

\AtBeginDocument{ 
    \fancypagestyle{firstpagestyle}{
        \fancyhf{}
        \fancyfoot[L]{\sffamily\footnotesize \workshopname}%
        \fancyfoot[C]{\sffamily\footnotesize \thepage}
    }
    \fancyhf{}
    \fancyhead[L]{\sffamily\footnotesize\shorttitle}
    \fancyhead[R]{\sffamily\footnotesize\shortauthors}
    \fancyfoot[L]{\sffamily\footnotesize\workshopname}%
    \fancyfoot[C]{\sffamily\footnotesize\thepage}
    \extrafootertext{\licensedetails}
}

\begin{document}

\title[Beyond the Winding Path of Learning]{Beyond the Winding Path of Learning: Exploring Affective, Cognitive, and Action-Oriented Prompts for Communication Skills}

\author{Naoko Hayashida}
\email{hayashida.naoko@fujitsu.com}
\orcid{0000-0002-6505-8364}
\affiliation{%
  \institution{Fujitsu Limited}
  \city{Kawasaki}
  \country{Japan}
}



\begin{teaserfigure}
\includegraphics[width=\textwidth]{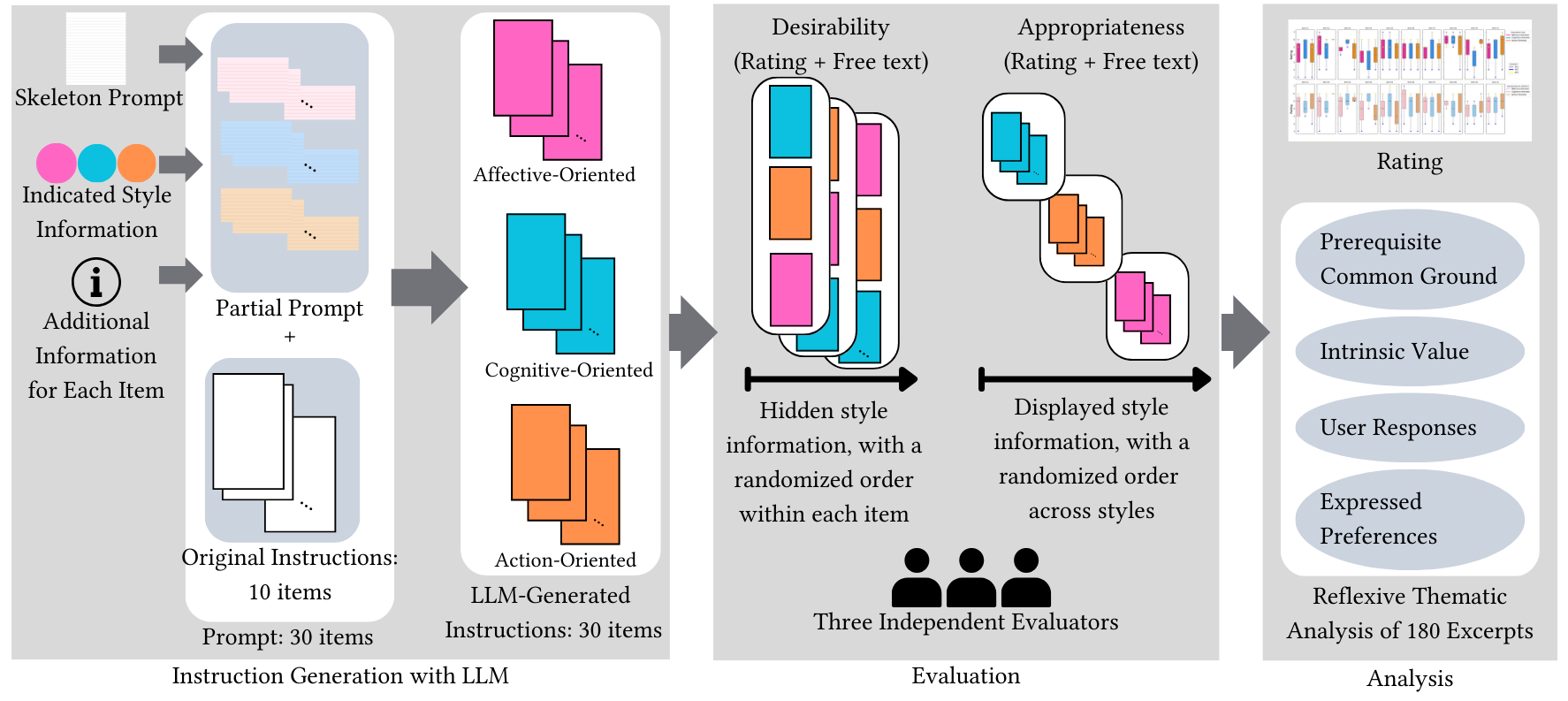}
\caption{Outline of the Study.}
\Description{LLM-generated instructions were evaluated for desirability and appropriateness by three independent evaluators, followed by Reflexive Thematic Analysis of 180 excerpts.}
\label{fig:fig01_Teaser} 
\end{teaserfigure}

\maketitle

\section{Introduction}
Since high dropout rates in online learning platforms were reported \cite{onah_dropout_2014}, various factors affecting learner retention have been identified \cite{aldowah_factors_2020}, with learners’ perceptions of their experiences playing a crucial role in shaping their persistence. For instance, Kittur et al. \cite{kittur_role_2021} highlight how success expectations are shaped by perceived system fit and course difficulty.

Recent advances in generative Artificial Intelligence (GenAI) present new possibilities for GenAI-mediated learning \cite{wang_llm-powered_2025, impey_using_2025}. AI-generated instructional messages are often perceived as clearer than human-written content\cite{lim_artificial_2023}, but their impact on learners’ perceptions of skill-building experiences remains underexplored.

This study examines GenAI-mediated learning in a self-directed context, focusing on communication skills. We compare three messaging styles—Affective, Cognitive, and Action-Oriented—to investigate their influence on learners’ perceptions of the learning process. We applied this approach to ten instructional units, using GenAI to generate 30 learning items. Three evaluators assessed them for desirability and appropriateness through numerical ratings and open-ended feedback. The 180 excerpts were analyzed using reflexive thematic analysis, revealing four overarching themes: Prerequisite Common Ground, Intrinsic Value, User Responses, and Expressed Preferences.

We discuss these insights to inform the design of GenAI-mediated, self-directed skill-building, with the goal of enhancing engagement, persistence, and learning outcomes.

\section{Related Work}
In the online teaching and learning research review, learners’ characteristics are classified by self-regulation, motivational, academic, affective, cognitive, and demographic factors \cite{martin_systematic_2020}. The characteristics of learners, including their expectations of course success, affect their persistence intentions \cite{kittur_role_2021}.  
As a learner-centered approach, rather than the traditional classroom model, learners’ preferences for content modality (such as VARK; V: visual, A: auditory, R: reading/writing, and K: kinesthetics) have been examined \cite{fleming_im_1995}.  
Regarding information value, Kelly and Sharot show that participants assess whether information is useful in directing action, how it will make them feel, and whether it relates to concepts they think of often \cite{kelly_individual_2021}.  
In our study, we examine how learners’ information preferences (i.e., their expectations and perceived value) should inform the design of GenAI-mediated self-study services.

\section{Design of Affective, Cognitive, and Action-Oriented Prompts Supporting Skill Development}

\begin{table}[h]
  \caption{Checkpoints for ensuring consistency in messaging styles.}
  \label{tab:tab01_checkpoints}
  \begin{tabular}{p{4cm} p{4.2cm} p{5.6cm}}
    \toprule
     Affective-Oriented & Cognitive-Oriented & Action-Oriented \\
    \midrule
    - Reassures the reader \newline
    - Considers the reader’s feelings \newline
    - Reflects broad endorsement \newline 
    - Respects autonomy (e.g., “Please try” instead of “Please do”) & 
    - Clarifies the learning item’s role within skill-building \newline
    - Emphasizes importance \newline
    - Highlights relevance to real life\newline
    - Backs up claims with scientific evidence & 
    - Provides concrete thinking strategies or practical methods \newline
    - Provides step-by-step guidance \newline
    - Uses examples from experienced individuals \newline
    - Helps learners visualize their own success \\
  \bottomrule
\end{tabular}
\end{table}

Each instructional text was generated using a Large Language Model (LLM), specifically ChatGPT o1 pro mode\footnote{ChatGPT o1 pro mode, \url{https://openai.com/index/introducing-chatgpt-pro/}}, based on 30 prompts.  
The prompt structure, as illustrated in Fig. \ref{fig:fig01_Teaser}, consists of four key components: 
\textbf{Skeleton Prompt}, a standardized template for instructional content; 
\textbf{Style Information}, which defines the messaging style (Affective, Cognitive, or Action-Oriented) and ensures consistency via predefined checkpoints (Table \ref{tab:tab01_checkpoints}); 
\textbf{Additional Information}, which includes style-specific details such as a warm introduction (Affective), research citations (Cognitive), or procedural guidance (Action-Oriented); and 
\textbf{Original Instructions}, a collection of 10 learning items (approximately 400 characters each in Japanese) that explain fundamental aspects of communication skills. The instructional content follows the principles of Nonviolent Communication (NVC)\footnote{NVC is based on the work of Marshall B. Rosenberg and the Center for Nonviolent Communication (\url{www.cnvc.org}).}, covering Observation, Emotion, Needs, and Requests.

\begin{figure}[h] 
\centering 
\includegraphics[width=0.98\linewidth]{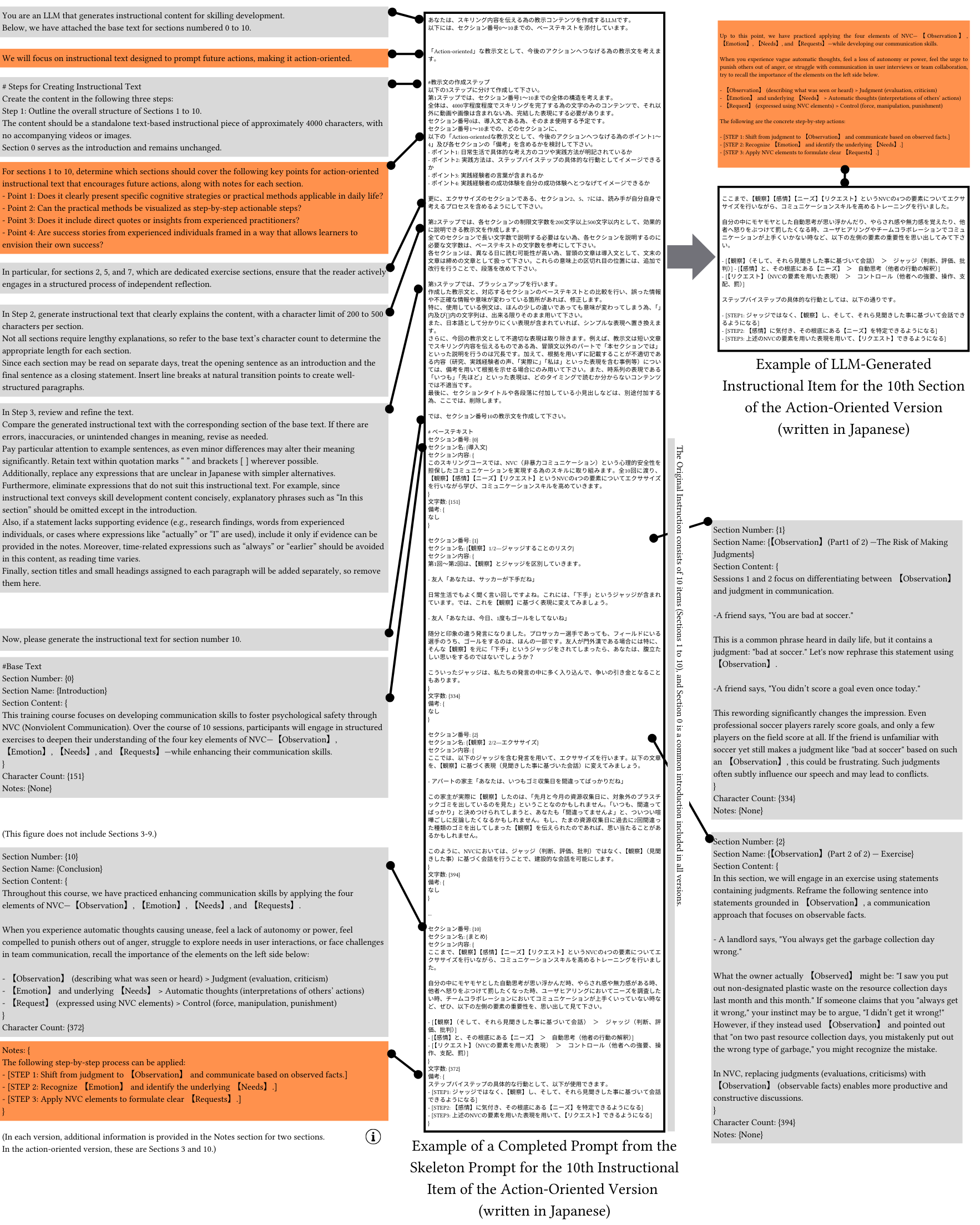} 
\noindent Note: Orange highlights denote Action-Oriented-specific elements.  
\caption{Example of a Completed Prompt and LLM-Generated Instruction Item.} 
\Description{This figure presents an example of a completed prompt derived from the skeleton prompt and the corresponding LLM-generated instruction item. The Original Instruction consists of 10 items (Sections 1 to 10), while Section 0 serves as a shared introduction across all versions. Additional information is included in the Notes section for two sections in each version. In the action-oriented version, these sections are Sections 3 and 10.}
\label{fig:figApdix01_Prompt} 
\end{figure}

A detailed breakdown of the prompts is shown in Fig. \ref{fig:figApdix01_Prompt}.

As shown in the Evaluation section of Fig. \ref{fig:fig01_Teaser}, three independent evaluators assessed the 30 instructional texts generated by the LLM twice. 
In the \textbf{First Evaluation(Desirability Assessment)}, style information was not disclosed; evaluators were presented with three instructional texts per learning item in randomized order and rated desirability using a 7-point Likert scale (1: Strongly Disagree – 7: Strongly Agree) along with open-ended feedback (e.g., aspects they found desirable or undesirable). 
In the \textbf{Second Evaluation (Appropriateness Assessment)}, style information was disclosed, and evaluators rated the appropriateness of texts within their assigned style using the same Likert scale and open-ended feedback. To minimize ordering effects, the sequence of style presentation varied across evaluators.

\begin{table}[t]
  \caption{Evaluators’ Characteristics and Perceived Necessity of Skilling.}
  \label{tab:tab02_evaluators}
  \begin{tabular}{llllll}
    \toprule
     ID & Age group  & Familiarity of NVC (pre) & Necessity (pre) & Necessity (post) & Necessity of NVC (post)\\
    \midrule
    e01 & 70s & 1 & 5 & 5 & 3\\
    e02 & 40s & 5 & 3 & 1 & 1\\
    e03 & 50s & 1 & 3 & 6 & 6\\
  \bottomrule
\end{tabular}
\\Note: This table summarizes evaluator characteristics, including their prior NVC familiarity and perceived importance of communication skilling before and after evaluation. Ratings were on a 7-point Likert scale (1: least, 7: most).
\end{table}

\begin{figure}[h] 
    \centering 
    \includegraphics[width=1.0\linewidth]{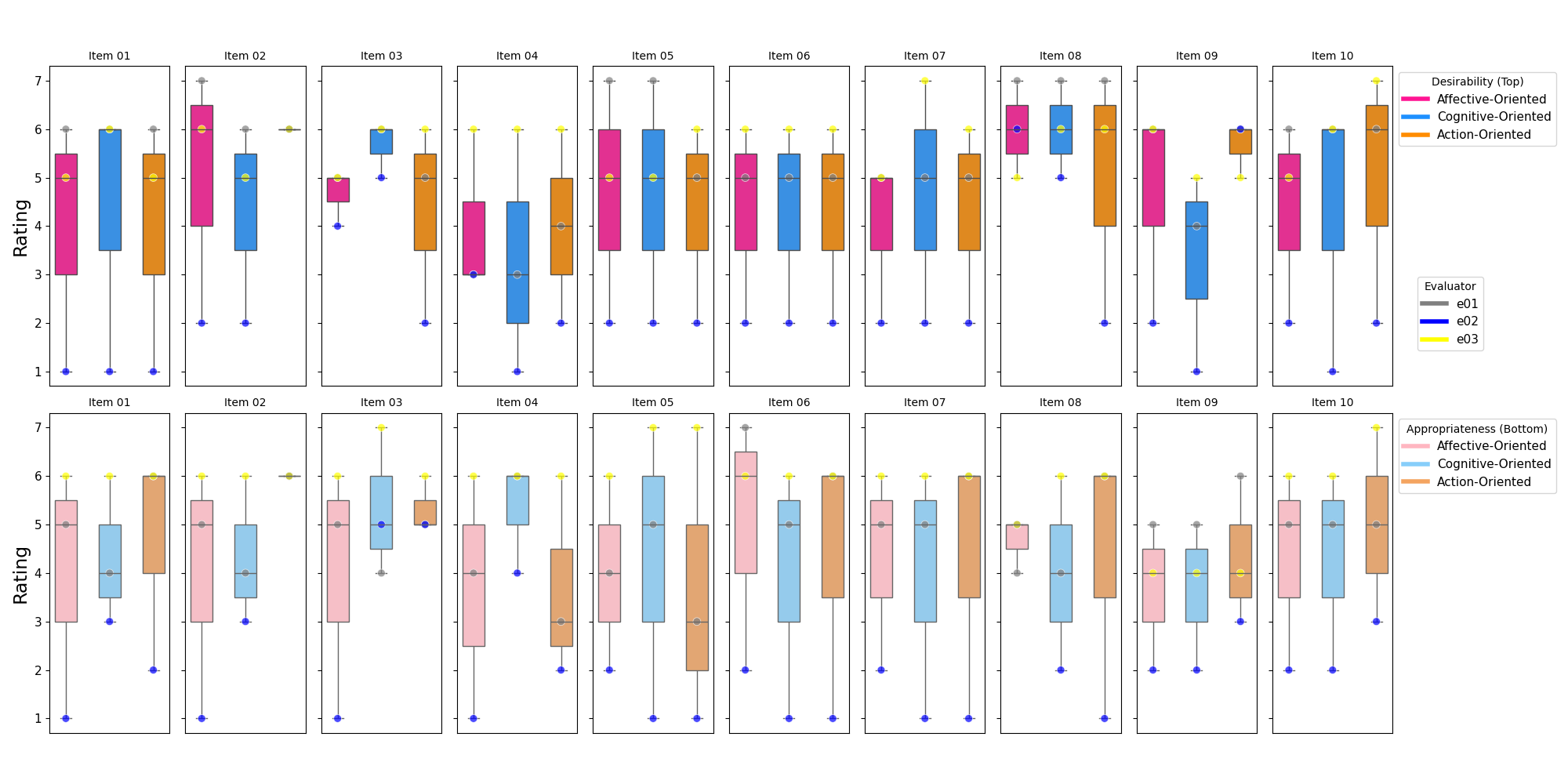} 
    \noindent Note: Ratings were on a 7-point Likert scale (1: least, 7: most).
    \caption{Evaluator Ratings for Desirability (Top) and Appropriateness (Bottom) of Instructional Items.} 
    \Description{This figure presents the numerical ratings of 10 instructional items, evaluated by three raters (e01, e02, e03) across two assessment phases: Desirability (Top) and Appropriateness (Bottom). Ratings are categorized by messaging style—Affective-Oriented, Cognitive-Oriented, and Action-Oriented—using distinct colors. Box plots represent rating distributions, with individual evaluator scores marked in gray (e01), blue (e02), and yellow (e03).}
    \label{fig:figApdix02_UserResponses} 
\end{figure}

Evaluators assessed the texts from the perspective of learners receiving LLM-generated skilling messages. Table \ref{tab:tab02_evaluators} summarizes their key characteristics, including perceived necessity of skill-building before and after the evaluation. The numerical results are presented in Fig. \ref{fig:figApdix02_UserResponses}.

In the next section, we conduct a reflexive thematic analysis \cite{braun_thematic_2012, braun_thematic_2021} of open-ended responses to explore learners’ perceptions in GenAI-mediated learning experiences.

\section{Insights on Learners’ perceptions}
We conducted a reflexive thematic analysis of 180 excerpts to explore learners’ perceptions in the communication skilling context. The analysis followed six steps: dataset familiarization, data coding, initial theme generation, theme development and review, theme refining, and defining and naming. Initial themes were derived by identifying properties and dimensions from one-third of the dataset, then refined iteratively. Table \ref{tab:tabApdix02_rtaresults} presents the extracted themes with representative comments, where excerpt IDs such as et001 indicate excerpt identification numbers.

We extracted themes related to \textbf{Prerequisite Common Ground(PCG)}, \textbf{Intrinsic Value(IV)}, \textbf{User Responses(UR)}, and \textbf{Expressed Preferences(EP)}.
We label each subtheme using a prefix notation: the theme acronym, an underscore, and the subtheme name (i.e., \textbf{EP\_Supporting Evidence} represents the “Supporting Evidence” subtheme under the EP).

\subsection{Observations}

\subsubsection{Explicit Preferences for Cognitive and Action-Oriented Content}
Evidence-based content such as research references (\textbf{EP\_Supporting Evidence}, et161), explanations facilitating actionable steps (\textbf{EP\_Actionable Instruction}, et150), and information-rich content (\textbf{EP\_Informative Content}, et103) were described as desirable.

\subsubsection{Self-Direction Fosters Positive and Engaged Responses}
Descriptions highlighted not only positive reactions driven by the perceived usefulness of the content (\textbf{UR\_Engaged Responses}, et009) but also the preference for content that allows room for autonomous thinking (\textbf{IV\_Self-directed Engagement}, et112).
If learners have strong preferences for the learning contents, the flexible learning paths can be one of the solutions to facilitate motivated feeling \cite{reinhard_one-size-fits-all_2024}.

\subsubsection{Negative Responses Arising from a Lack of Shared Understanding}
Negative reactions were observed when shared understanding was not established. This included expressing opinions about the characters’ behaviors in the learning content (\textbf{UR\_Varied Low-Engagement Responses (Opinion Sharing without Avoiding Learning)}, et023) and showing negative responses to the learning content itself (\textbf{UR\_Varied Low-Engagement Responses (Indicating Learning Avoidance)}, et003).
For example, evaluator e01 initially resisted engaging with content, stating, \textit{“...expressing emotions that cause stress is undesirable...”} (et003). However, by et009, after progressing through the content, their response became more positive. This suggests that when self-awareness-based reflection, such as organizing emotions within feelings, is absent, learners may initially rely on pre-existing cognitive frameworks, leading to misinterpretation.

Additionally, while it is well known that a fixed mindset—the belief that skills are static and unchangeable—can hinder learning\cite{dweck_mindset_2006}, our data suggests that the inability to establish a shared conceptual framework (\textbf{PCG\_Shared Conceptual Framework}, et003) precedes this resistance. This lack of shared understanding then leads to a stronger rejection, manifesting as beliefs such as “skills cannot change” or “I don’t want to go that far” (\textbf{PCG\_Fundamental Mindset}, et025). Furthermore, some learners did not recognize that LLM-generated instructional texts employed strategies to facilitate cognitive, affective, and action-oriented learning (\textbf{PCG\_System Understanding and Acceptance}, et091).

The role of “recipient design” is crucial in communication, as noted by Mustajoki\cite{mustajoki_speaker-oriented_2012}.
In our study, we observed that users who failed to establish common ground with the agent exhibited negative reactions. Tolzin and Janson \cite{tolzin_mechanisms_2023} identify key mechanisms for achieving common ground in human-agent interactions, including embodiment, social features, joint action, knowledge base, and the mental model of conversational agents. These mechanisms emphasize the importance of the agent’s social features and knowledge base in fostering effective communication. Building upon this, it is also crucial to establish common ground concerning the content of the skilling material\cite{azevedo_scaffolding_2005}. Aligning the agent’s knowledge, including instructional content and social attributes, with the user’s expectations can enhance user engagement and facilitate more effective learning experiences.

\subsection{Open Questions for the Workshop}

\subsubsection{How Should We Design to Avoid Overly Affective Content While Maintaining a Frictionless Experience?}
Regarding verbal expressions, an appropriate level of politeness without excess was preferred (\textbf{EP\_Politeness in Expression}, et130). Additionally, maintaining a frictionless reading experience was emphasized to ensure continued engagement (\textbf{IV\_Positive and Frictionless Learning Experiences}, et130).

Our study did not reveal a strong preference for verbal expressions intending affectively oriented instruction. However, in systems where learners initiate inquiries and receive system responses—such as AutoTutor \cite{graesser_autotutor_2005}—incorporating affective elements like empathetic listening \cite{arjmand_empathic_2024} may foster more positive learning experiences. AutoTutor engages learners through mixed-initiative dialogue, adapting dynamically based on cognitive and affective cues, which has been shown to enhance engagement and learning outcomes. Similarly, research on empathetic system responses suggests that fostering affective connections can help sustain motivation and deepen interaction.

These insights indicate that while overly affective content may be undesirable, embedding affective components into system interactions rather than the instructional text itself could enhance engagement and create a more supportive learning environment.

\subsubsection{Do People Express Their Opinions to an LLM as They Would Do to Humans?}
During evaluations, evaluators attempted to share personal values (\textbf{IV\_Desire to Share Personal Values}), such as \textit{“...I believe it is undesirable to have negative emotions”} (et023), suggesting that when learners express their opinions, human instructors can identify and address potential misunderstandings of the instructional content. However, when users disclosed messages generated by AI, their attitudes differed from when interacting with humans \cite{lim_effect_2024}. This difference is not necessarily negative, but it highlights the importance of considering whether the actor is AI or human, as this distinction influences user behavior. Therefore, it is crucial to design interactions that account for what types of information users feel comfortable sharing with LLMs, ensuring a motivating and supportive learning environment.

\begin{acks}
We thank the anonymous reviewers and all those who participated in this project for their useful comments and help.
\end{acks}

\bibliographystyle{ACM-Reference-Format}
\bibliography{zotero_bibtex20250214}

\appendix
\counterwithin{figure}{section}
\counterwithin{table}{section}

\section{Supplementary Materials (Figures \& Table)}

\begin{table}[h]
    \caption{Reflexive Thematic Analysis Results}
    \label{tab:tabApdix02_rtaresults}
    \begin{tabular}{p{2cm} p{4cm} p{8.5cm}}
        \toprule
        \textbf{Theme} & \textbf{Sub-Theme} & \textbf{Example excerpt} \\
        \midrule
        \textbf{Prerequisite Common Ground\newline(PCG)} & \textbf{PCG\_Shared Conceptual Framework} & \textit{“It is desirable to express one’s emotions, and as we get older, emotions tend to fade. However, expressing emotions that cause stress is undesirable and, I believe, not good for health.” } et003 (e01\_1st07\_Affective)\\
                         & \textbf{PCG\_System Understanding and Acceptance} & \textit{“In the first place, what exactly emotional connection refers to is ambiguous. If it means communicating without making the other person uncomfortable, I do not think that the rephrasing in this example can create an emotional connection.” } et091 (e02\_2nd11\_Affective)\\
                         & \textbf{PCG\_Fundamental Mindset} & \textit{“I think it is desirable to observe what happens around me. However, I do not wish to go as far as writing down emotions in concrete words.” } et025 (e01\_1st15\_Action)\\
        \midrule
        \textbf{Intrinsic Value(IV)} & \textbf{IV\_Positive and Frictionless Learning Experiences} & \textit{“The opening part, ‘I sincerely appreciate you for reading this far. Even if you felt uncertain or confused during the learning process, I would be glad if you could read on with even a little sense of reassurance,’ gives a somewhat overly polite impression. It makes me feel less inclined to continue reading.” } et130 (e03\_1st29\_Affective)\\
                         & \textbf{IV\_Self-directed Engagement} & \textit{“The lack of specific examples of observation is actually a good thing, as it does not restrict the readers’ free thinking.” } et112 (e02\_2nd02\_Action)\\
                         & \textbf{IV\_Desire to Share Personal Values} & et003, et025, et023 \\
        \midrule
        \textbf{User \newline Responses\newline(UR)} & \textbf{UR\_Varied Low-Engagement Responses (Opinion Sharing without Avoiding Learning)} & \textit{“I think it is desirable to control emotions under stress. I believe it is undesirable to have negative emotions.” } et023 (e01\_1st08\_Action)\\
                         & \textbf{UR\_Varied Low-Engagement Responses (Indicating Learning Avoidance)} & et003, et025\\
                         & \textbf{UR\_Engaged Responses} & \textit{“
                         Not only my team but also I myself sometimes lack the process of observation, emotions, needs, and requests. I think it is desirable to take a step back and reflect calmly.” } et009 (e01\_1st27\_Affective)\\
        \midrule
        \textbf{Expressed Preferences(EP)} & \textbf{EP\_Politeness in Expression} & et130 \\
                         & \textbf{EP\_Supporting Evidence} & \textit{“Specific evidence from psychological research is clearly presented, making the content intellectually engaging.” } et161 (e03\_1st01\_Cognitive)\\
                         & \textbf{EP\_Actionable Instruction} & \textit{“I found it good that specific actions were introduced at the end, compared to other patterns.” } et150 (e03\_1st30\_Action)\\
                         & \textbf{EP\_Informative Content} & \textit{“Since the technical term ‘automatic thoughts’ appears, those who are interested may find their intellectual curiosity stimulated.” } et103 (e02\_2nd23\_Cognitive)\\
        \bottomrule
    \end{tabular}
    \\Note: Excerpt IDs, such as et001 (e01\_1st01\_Affective), consist of an excerpt identification number (et001), evaluator identification number (e01), evaluation round (1st or 2nd), evaluation order within that round (01–30), and messaging style (Affective, Cognitive, or Action).
\end{table}

\end{document}